\documentclass[%
reprint,
superscriptaddress,
 amsmath,amssymb,
 aps,
pre,
]{revtex4-1}

\usepackage{graphicx}% Include figure files
\usepackage{dcolumn}% Align table columns on decimal point
\usepackage{bm}% bold math
\usepackage{hyperref}% add hypertext capabilities
\begin{document}

\preprint{}

\title{Experimental investigation of the stability of the floating water bridge}% Force line breaks with \\
%\thanks{A footnote to the article title}%

\author{Reza Montazeri Namin}
\affiliation{Department of Mechanical Engineering, Sharif University of Technology, Tehran, Iran.}%
\email{namin@mech.sharif.edu}
\homepage{http://mech.sharif.edu/~namin}

\author{Ahmad Amjadi}
\affiliation{Medical Physics Laboratory, Department of Physics, Sharif University of Technology, Tehran, Iran.}%

\author{Shiva Azizpour Lindi}%
\affiliation{National Organization for Development of Exceptional Talents (NODET), Farzanegan Highschool, Tehran, Iran.}%

\author{Nima Jafari}

\author{Peyman Irajizad}
\affiliation{Medical Physics Laboratory, Department of Physics, Sharif University of Technology, Tehran, Iran.}%

\date{\today}% It is always \today, today,
             %  but any date may be explicitly specified

\begin{abstract}
When a high voltage is applied between two beakers filled with deionized water, a bridge
of water may be formed in between exceeding the length of 2 cm when the beakers are
pulled apart.
We construct experiments in which the geometry and the electric field within the bridge
are measured and compared with predictions of theories on the floating water bridge.
A numerical simulation is used for the measurement of the electric field.
Our experimental results approve that two forces of dielectric tension and surface tension
are holding the bridge against gravity. These forces have the same order of magnitude. Results show
that the stability can be explained by macroscopic forces, regardless of the microscopic 
changes in water structure.

\begin{description}
%\item[Usage]
%Secondary publications and information retrieval purposes.

\item[PACS numbers]
%May be entered using the \verb+\pacs{#1}+ command. 

%\item[Structure]
%You may use the \texttt{description} environment to structure your abstract;
%use the optional argument of the \verb+\item+ command to give the category of each item. 
\end{description}
\end{abstract}
\maketitle

%\tableofcontents

\section{\label{sec:Intro}Introduction}

\begin{figure}[b]
 \includegraphics[width=\linewidth]{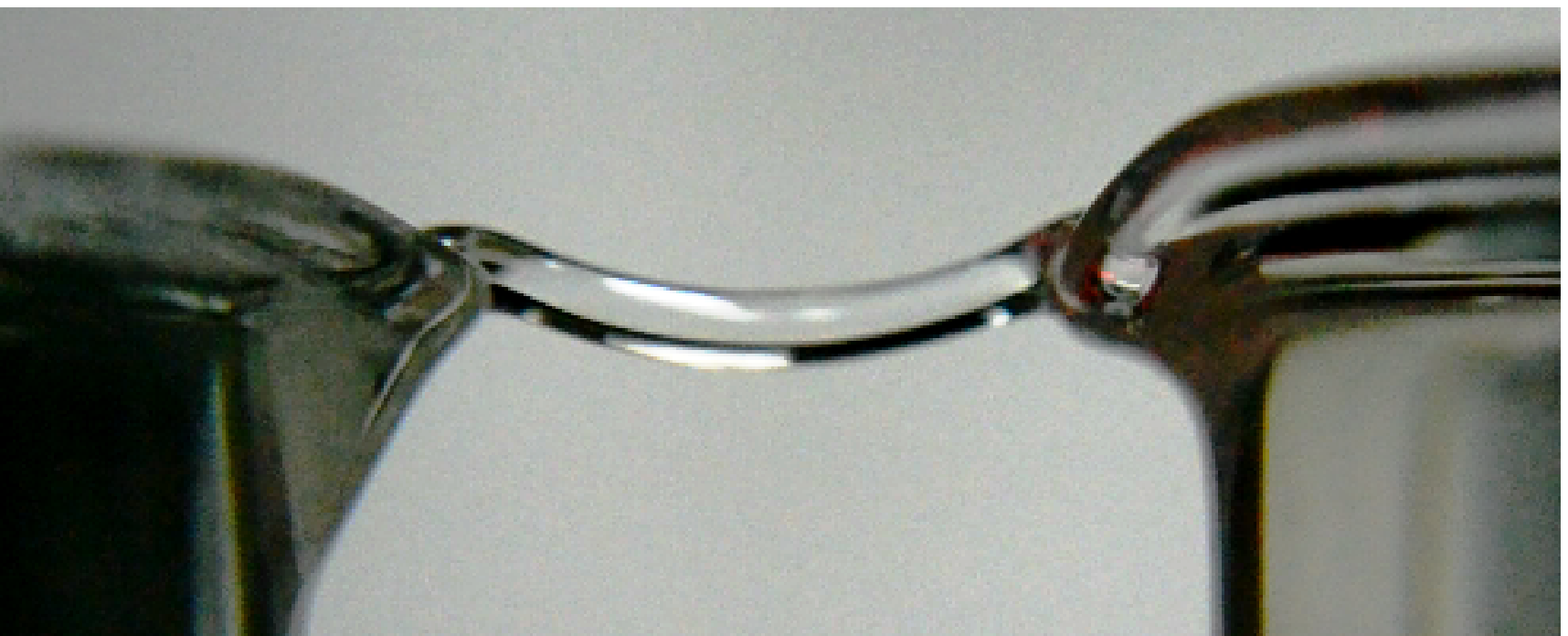}
 \includegraphics[width=\linewidth]{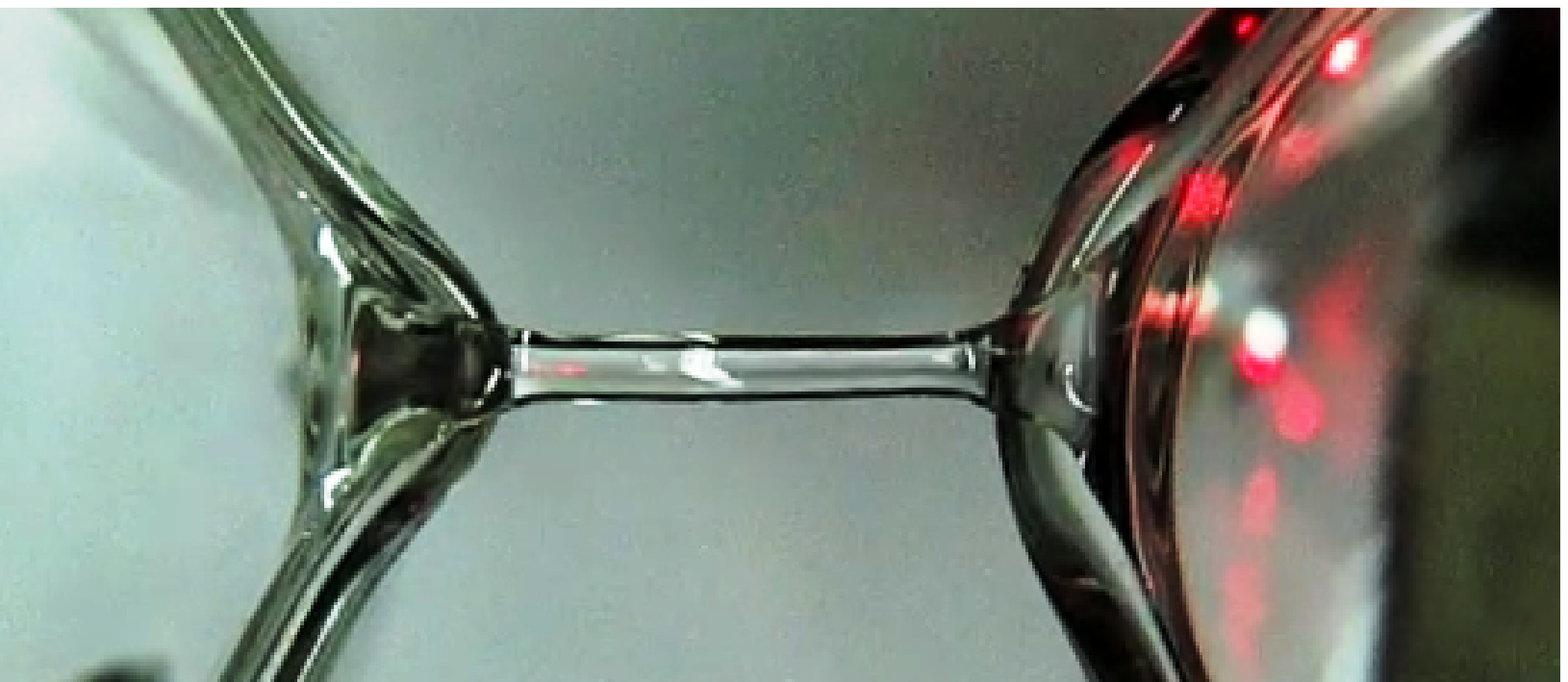}
 \caption{The floating water bridge from Front and Top View}\label{fig:pic}
\end{figure}

The floating water bridge is an interesting phenomenon first reported
by Armstrong in 1893 \cite{armstrong1893electrical}.After more than a century Fuchs et al
\cite{fuchs2007floating} reported their investigation about this interesting phenomenon
and showing the different behaviours in it, suggested that it could reveal some hidden
properties of water \cite{fuchs_dynamics}.
Two beakers filled with deionized water are subjected to a DC high voltage more than 10kV 
and a bridge is formed between them (figure \ref{fig:pic}) which can last for hours and have a 
length exceeding $2cm$ \cite{wois_exp}.
This experiment is stable, easy to reproduce and leads to a special condition that 
the water in the bridge can be accessed and experimented under high voltages and 
different atmospheric conditions \cite{fuchs2010can}. This has led to several special 
experiments in this setup, including Neutron scattering \cite{fuchs2009neutron, fuchs2010two}, 
visualization using optical measurement techniques \cite{fuchs_dynamics, wois_exp}, 
Raman scattering \cite{ponterio2010raman}, Brillouin scattering \cite{fuchs2009inelastic} 
and zero gravity experiments \cite{fuchs2010behaviour}, many of which have attempted to investigate
the possible structural changes in the water bridge causing its formation, stability and other properties observed
in the water bridge. Some observations have been explained by quantum electro dynamic theories \cite{del2010collective}.
Aqueous solutions have been tested under same setup and liquid bridging has been observed
and conductivity and mass transfer differences have
been investigated \cite{eisenhut2011aqueous} as well as thermal differences in the behaviour
of the bridge \cite{fuchs2011methanol}.
Also the bridging has been observed in liquids other than water \cite{marin_t}. Mid-infrared emission
investigation of the water bridge suggest the existence of micro and nano droplets electrosprayed from
the liquid-gas interface \cite{fuchs2012investigation}. Transport and behaviour of bacterial cells added
to the water bridge have also been investigated \cite{paulitsch2012prokaryotic}.
 
Reviews on this topic have been published \cite{fuchs2010can, woisetschlager2012horizontal} 
which the reader may refer to for a comprehensive literature review.

While many discussions have been published regarding the structural changes in water
leading to the stability of the bridge and suggest the existence of anisotropic chains
of molecules in the bridge, high energy X-ray diffraction experiments show no preferred
orientation of the molecules in the water bridge, which is also approved in molecular dynamics
simulations \cite{skinner2012structure}.
In the present investigation, we concentrate on the theories based on macroscopic forces
explaining the stability of the bridge.
In this case there are two different perspectives:
Widom et al in 2009 \cite{widom_t} suggest the existence of a tension along the bridge caused by
the electric field within the dielectric material. They provide theoretical calculations based
on the Maxwell pressure tensor within the dielectric to calculate this tension. The tension along
the curved water bridge causes an upward force defying gravity. Marin and Lohse 2010 \cite{marin_t} apply
a similar theory while the tension is calculated as half the value derived by \cite{widom_t}, and in a modified
experimental setup compare the results with experiments. They also measure water flow along the bridge
suggesting electrical charges responsible for that. Morawetz 2012 \cite{morawetz_t, morawetz2012effect} discusses the effect
of electrical charges in a charged catenary and solves the flow and derives stability criteria.

On the other hand, Aerov in 2011 \cite{aerov_t} claims to prove that the tension caused by electric field
within the dielectric material is zero and the only force holding the bridge against gravity is surface
tension. The effect of the electric field according to Aerov is to avoid the breakup of the bridge into small
droplets and maintain stability.

We try to examine the theoretical perspectives experimentally by designing quantitative experiments
which are comparable with the two theories. The experimental setup is explained in section \ref{sec:Exp}.
We precisely measure the geometry of the bridge by image processing, and estimate the electric field with
use of a numerical simulation explained in part \ref{sec:EEval}. Experimental results and the theories
are compared in section \ref{sec:Res} leading to conclusion.

\section{\label{sec:Exp}Experiments}
The experimental setup consists of two $50ml$ beakers filled with deionized water.
The water was produced with a Millipore SimPak1 purification pack kit and initially
had a resistivity of 18.2  M$\Omega.$cm.
Resistivity of pure DI water decreases rapidly by contamination of impurities, e.g. the
$CO_2$ gas from air. The resistivity in our experiments was reduced to 1.8 M$\Omega.$cm.
The resistivity also varies by temperature changes.It decreases from 1.8 M$\Omega.$cm at
25$^\circ$C to 1 M$\Omega.$cm at 45$^\circ$C.

\begin{figure}
 \includegraphics[width=\linewidth]{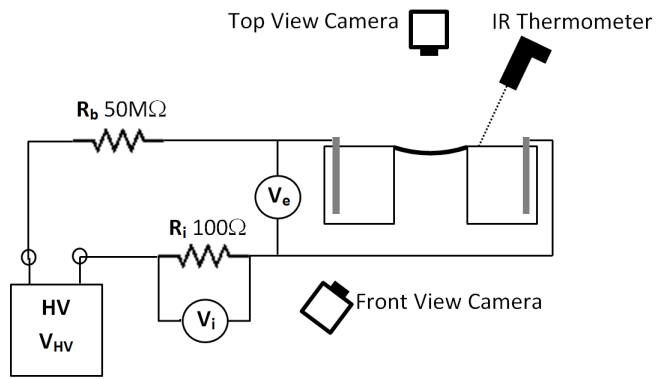}
 \caption{\label{fig:setup} Schematic of the experimental setup.}
 \end{figure}

A high voltage power supply was used which could provide a voltage upto 25 kV with 20 mA
current intensity (Plastic Capacitors HV250-103M). A resistance of 50 M$\Omega$ was placed 
after the power supply as a ballast resistor to control the current in the
circuit which had a great effect on the stability of the bridge, as shown in figure \ref{fig:setup}.
Also a resistance
of 100 $\Omega$ was placed so that the voltage difference along it demonstrates the current intensity.
Two aluminium plates were placed at the far ends of the beakers in the water connected to
the power supply. The voltage difference between the electrodes and the current intensity
were measured.
An infra-red thermometer (TES 1326S) was used to measure the surface
temperature of water. The current intensity and experimenting time were kept small enough
so that the temperature rise of the surface of water did not exceed 2$^\circ$C
and was kept between 24$^\circ$C and 26$^\circ$C during the experiments. So there might be about $5\%$ of change 
of resistivity in different points in water.

Two cameras were recording the bridge from the top view and the front view. 
A third camera was recording the current intensity and voltage.
For the extraction of quantitative data, we developed an image processing code using MATLAB to
read the three movies and extract the desired data which includes the average diameter from the top
view $D_t$, the average diameter from the front view $D_f$, the curvature of the centreline of the bridge at
it's centre $\xi$, the voltage $V$ and the current intensity $I$ at every frame. To estimate the curvature of the
centreline,initially the centreline was calculated by averaging the top and bottom of the bridge from the front
view, then a parabola was fitted to the line and using the coefficient of the parabola the curvature was
estimated. Also an error to this curvature was estimated by the regression of the fitted curve. The amount of
voltage was used to estimate the electric field in the centre of the bridge as explained in section \ref{sec:EEval}.

\section{\label{sec:EEval}Evaluation of the Electric field}
To compare the shape of the bridge in experiments with the predictions of the theories, the precise value
of the electric field was to be measured experimentally. For this propose we
used a numerical simulation; the electric current (ec) module of the COMSOL Multiphysics program was used.
The geometry of the bridge was approximately modelled and the average electric field was calculated
at a cross section in the middle of the bridge ($E$).

In the simulation, water was assumed to be a conducting material with a constant resistivity and 
dielectric permittivity. The shape of the bridge and water in the beakers was carefully estimated
by measuring the geometry of the beaker tips. During the experiments
the beakers were filled with water to the top, so that the unknown geometry of the water connecting the 
water in the beaker to the water in the bridge becomes least important.

 \begin{figure}
 \includegraphics[scale = 0.15]{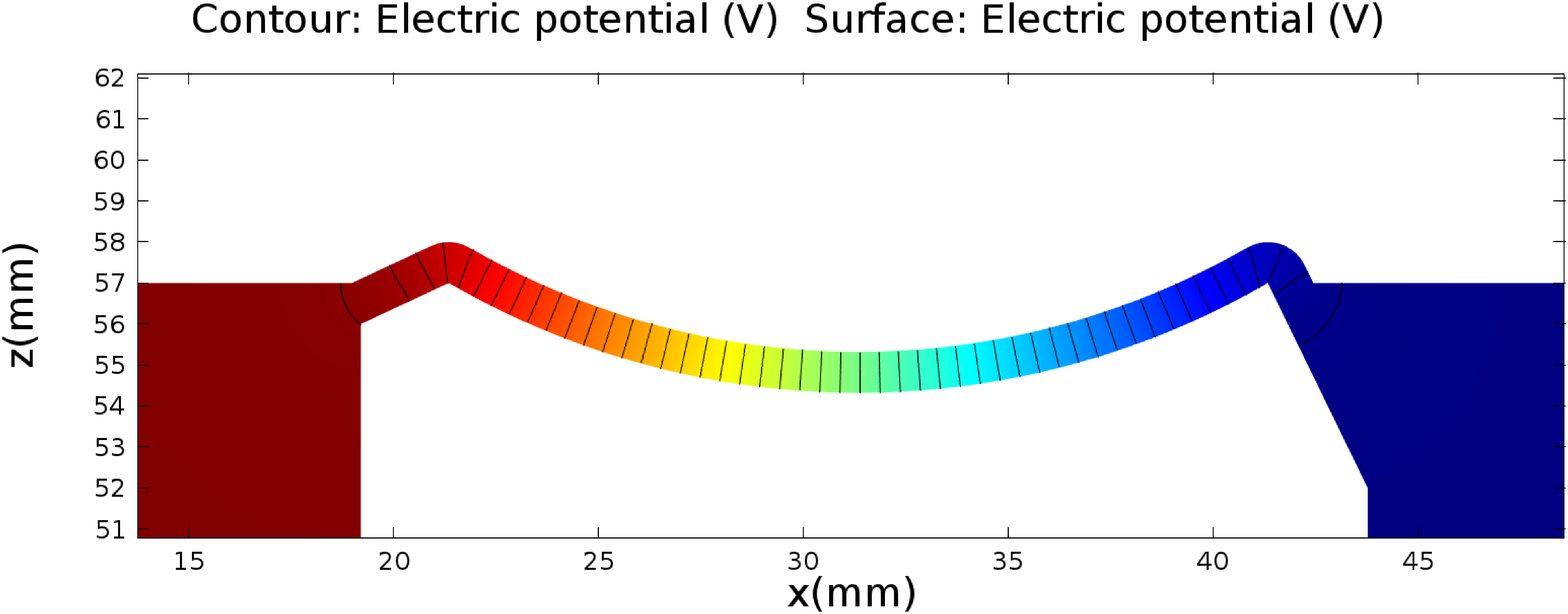}
 \includegraphics[scale = 0.15]{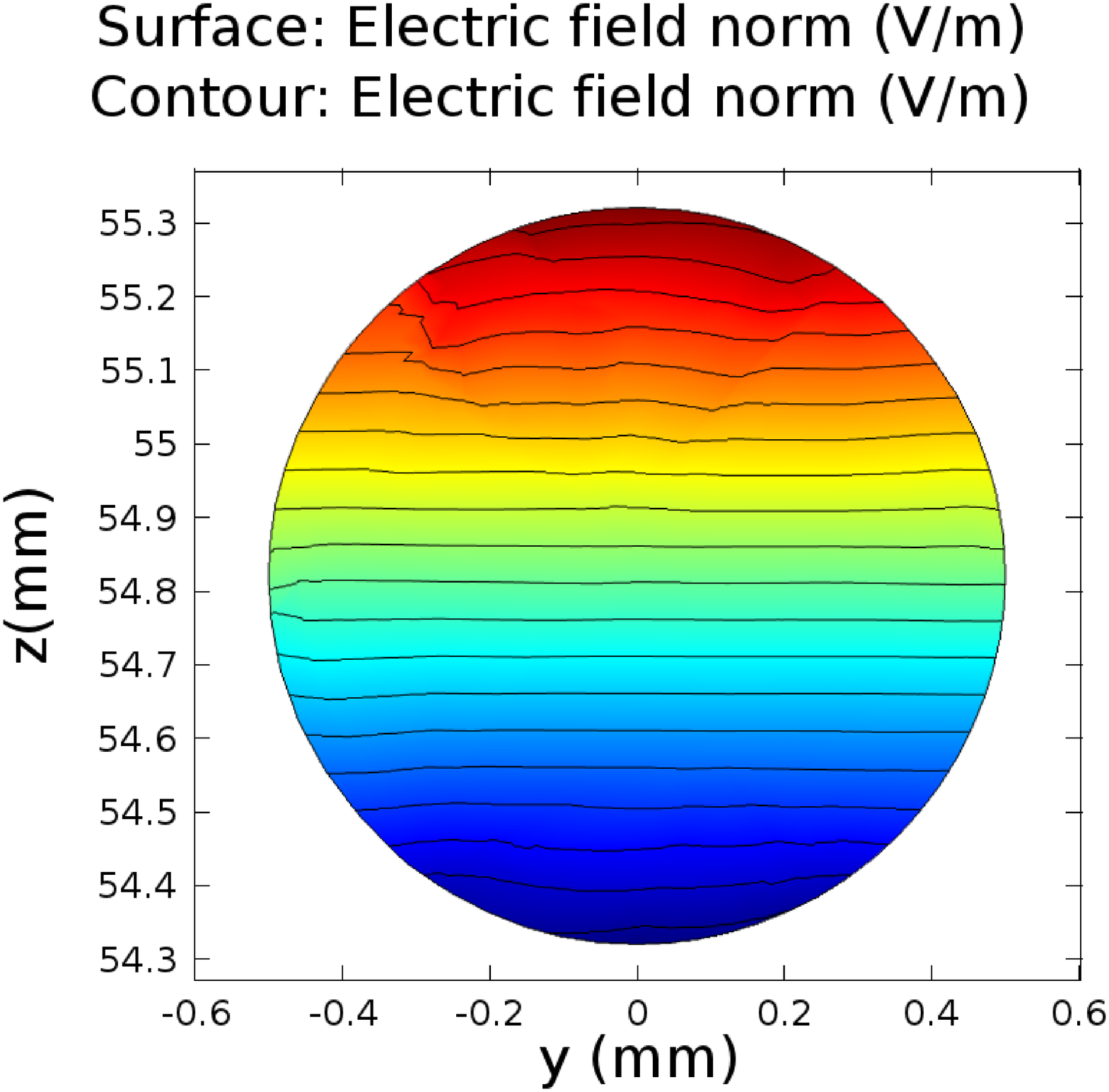}
 \includegraphics[scale = 0.15]{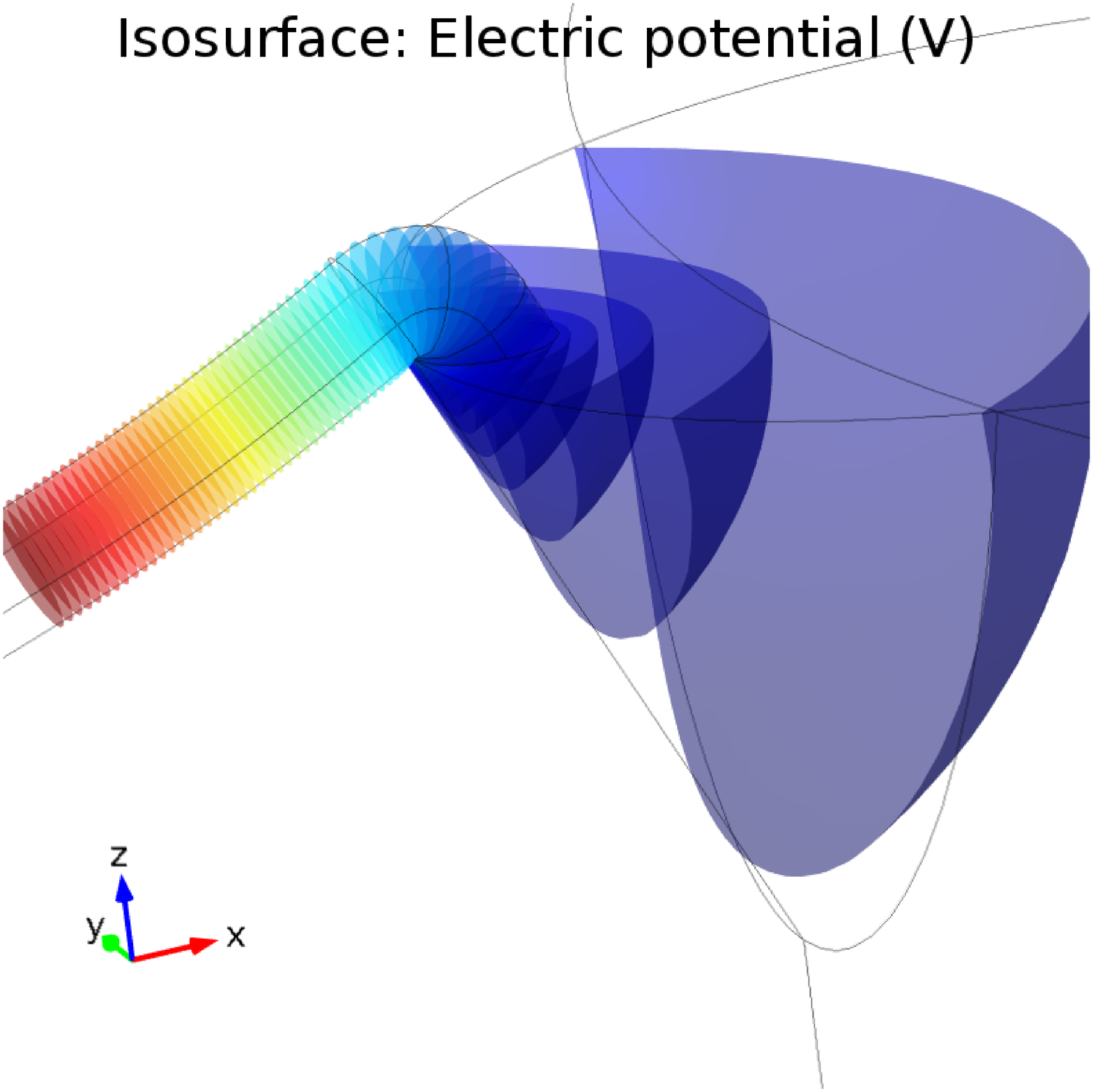}
 \caption{\label{fig:num1}Results of the Numerical Simulation.
 a) Contours of Electric Potential on a Cut Plane at $y = 0$. It shows that the equipotential surfaces are oriented normal to the bridge centreline.
 b) Contours of Electric Field Intensity on a Cross-Section at the Middle of the Bridge. The upward gradient of the electric field is the cause
 of the dielectric force holding the bridge. Calculations regarding this factor are presented in section \ref{sec:Res} and Appendix.
 c) Equipotential Surfaces Near the Tip of One Beaker.}
 \end{figure}

\begin{figure}
 \includegraphics[width=\linewidth]{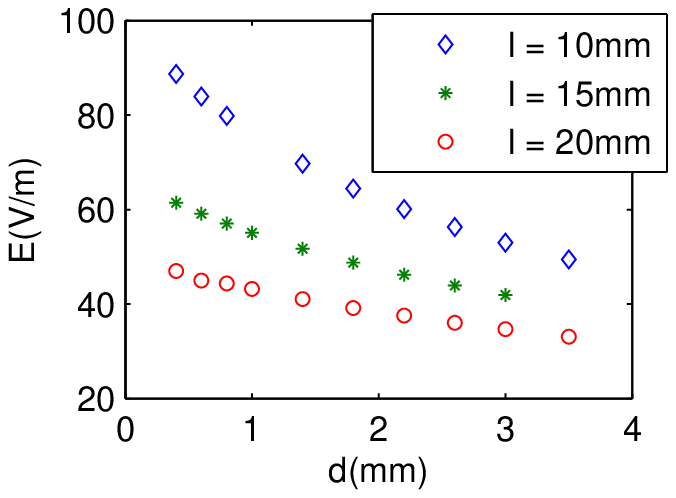}
 \includegraphics[width=\linewidth]{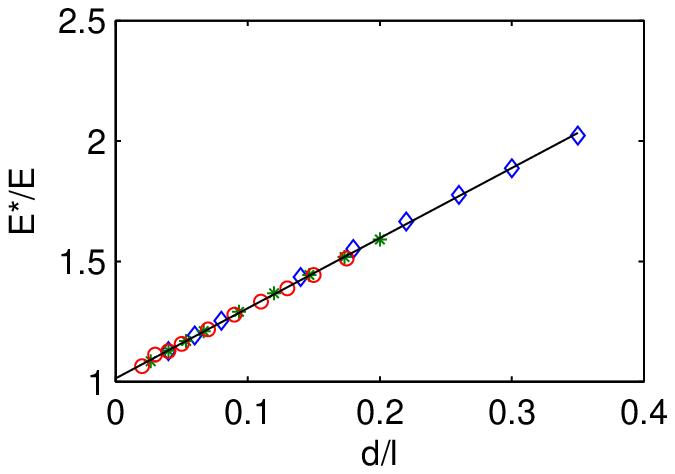}
 \caption{\label{fig:num_res1} a) Evaluated Electric Field From the Numerical Simulation.
$V = 1V$, $\xi = 0.001$, for three values of $l$.
 b) Analysed Data. The line represents a linear fit: 
 $E^*/E = 2.9096 (D/l) + 1.0144$ with the Regression of $R^2=0.98$.
 }
 \end{figure}

As a result of the simulation, it was observed that the electric field decreases 
with the increase of diameter of the bridge.
Also defining $E^* = V / l$, the ratio $E / E^*$ increases and approaches to one 
with the increase of the bridge length.
This ratio seems to be only a function of $l/D$, and not a function of $l$ and $D$ independently. 
According to figure \ref{fig:num_res1} b),
there is a linear relation between $E^* / E$ and $D/l$. 
The effect of curvature on the electric field was less than $1.5\%$ in the range of
experiments and was neglected. Also an estimation on the uncertain geometric properties was preformed, 
such as the elevation difference
of the water in the beaker (which in experiments was less than $0.5mm$), and as a result 
errors caused by this geometric approximations
was less than $8\%$ for $l=10mm$ and less than $5\%$ for $l = 20mm$. This was the most 
important factor in determining the error bars.

\section{\label{sec:Res}Results and discussion}
\begin{figure}
 \includegraphics[width=0.98\linewidth]{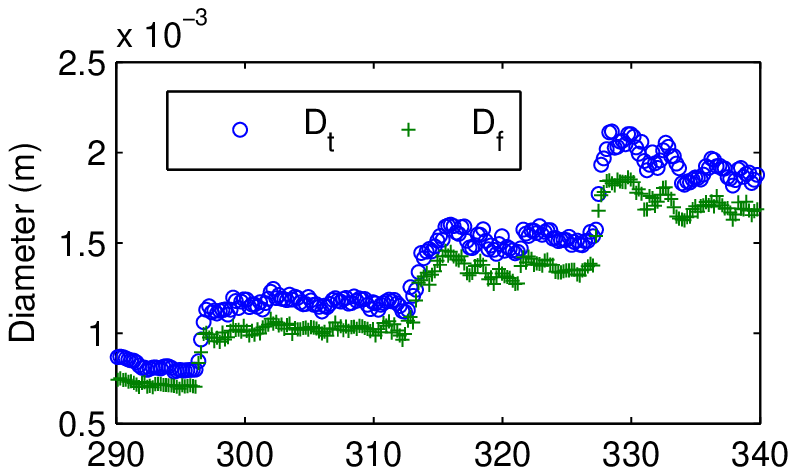}
 \includegraphics[width=0.98\linewidth]{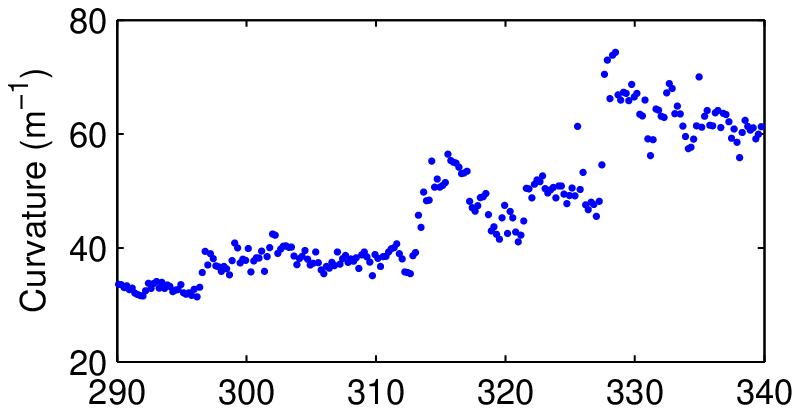}
 \includegraphics[width=0.98\linewidth]{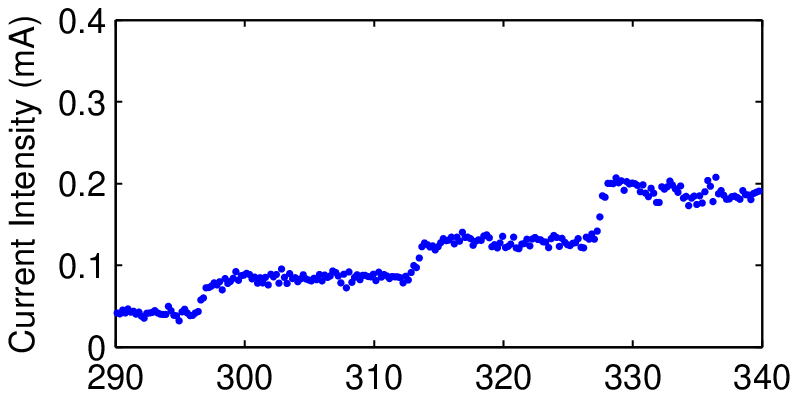}
 \includegraphics[width=0.98\linewidth]{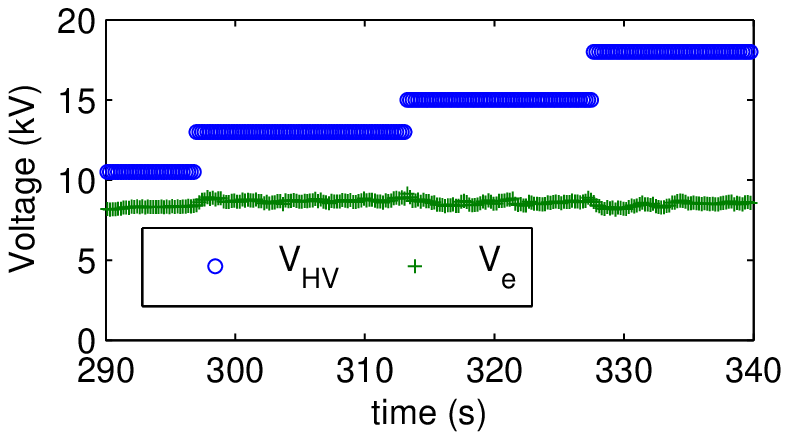}
 \caption{\label{fig:exp}Data extracted directly from experiments in means of time. 
 a) Diameters from top and side view
 b) Curvature of the bridge centreline 
 c) Current intensity passing through the bridge.
 d) Voltage differences across the electrodes ($V_e$) and the 
 power supply ($V_{HV}$). Time is started wit the bridge formation.}
 \end{figure}

The extracted data from experiments are the average diameter from top view $D_t$, average diameter from
front view $D_f$, curvature of the bridge from the front view $\xi$, voltage difference between the 
electrodes $V_e$ and current intensity $I$. Every five extracted data ($0.2s$) has been averaged to present one
data point. The distance between the beaker tips $l_b$ and the voltage
difference across the high voltage supply $V_{HV}$ were the independent parameters we could change 
during experiments. Our experimental data is presented in figure \ref{fig:exp}.

Figure \ref{fig:exp} shows that by increasing the voltage produced by the power supply, the voltage difference
between the electrodes does not change significantly. Instead, current intensity increases and the residual
voltage will be dropped at the ballast resistor. The reason for this fact seems to be that the diameter
of the bridge increases with the increase of current intensity, causing a fairly linear relation between current
intensity and cross-sectional area of the bridge. This causes the electric resistance of the bridge to drop
with an increase in current intensity and as a result the voltage drop across the bridge remains fairly constant.
In the absence of the ballast resistor, the water in the beaker acts fairly similar to the ballast resistor keeping the
voltage difference across the bridge constant in different current intensities.
We suggest this to be a reason for not achieving longer bridges in higher electric voltages; i.e. by
increasing the electric voltage, bridge increases its thickness passing a higher current intensity and the
electric field along the bridge remains constant and does not increase.
 
To analyse the experimental data, the average diameter of the bridge was calculated as $D = \sqrt{D_t D_f}$
and the electric field was estimated using the results of section \ref{sec:EEval} as a function of $V_e$ and
$D$ and the length of the bridge $l$.

To quantitatively compare our experiments with theoretical results of Widom et al \cite{widom_t}
and Aerov \cite{aerov_t}, their theory is used to find the fraction of their suggested forces
to the force needed to hold the bridge in experiments. In the equilibrium condition, since the
sum of the vertical forces holding the bridge should be zero, the total fraction of the holding
forces to the gravitational force must be equal to one.

The tension because of the electric field
in a dielectric medium calculated by Widom et al \cite{widom_t} follows this relation:
\begin{eqnarray}
T_{DE} = \varepsilon_0 (\varepsilon - 1) E^2 A
\label{eq:one}.
\end{eqnarray}
Where $A$ is the cross sectional area of the bridge, and is equal to $\pi D^2 /4$, $\varepsilon$ is the
relative permittivity of the water, which was assumed to be 80 and $\varepsilon_0$ is the vacuum permittivity.
If a tension $T$ is acting on a curved bridge with a curvature of $\xi$, the vertical force it causes
per unit length of the bridge is $\xi T$, while the gravitational force per unit length is $\rho A g$.
Thus the ratio of the dielectric force and gravitational force ($R_{DE}$) will be:
\begin{eqnarray}
%\xi = \dfrac{2\rho g}{\epsilon E ^ 2}
%\xi = \dfrac{\rho g}{\varepsilon_0 (\varepsilon - 1) E ^ 2}
R_{DE} = \dfrac{\varepsilon_0 (\varepsilon - 1) E ^ 2 \xi}{\rho g}
\label{eq:two}.
\end{eqnarray}
We have also calculated the upward force exerted by the electric field in a curved dielectric bridge
with a different method from Widom et al \cite{widom_t} which leads to the same force as they have calculated; explained in Appendix.
 
Aerov \cite{aerov_t} states that the electric tension along the bridge is zero, and the tension holding the 
bridge is the surface tension. The electric field causes the stability of the bridge and avoids
it's breakup to droplets. The tension caused by surface tension is the sum of the tension on
the sides ($ \gamma  P$) and the repulsing tension caused by pressure jump at the surface
($-\gamma P / 2$):

\begin{eqnarray}
T_{ST} = \dfrac{1}{2} \gamma P
\label{eq:three}.
\end{eqnarray}
where $P$ is the perimeter of the cross-section of the bridge and is equal to $\pi D$.
According to this assumption, the ratio between the upwards surface tension force and
gravitational force ($R_{ST}$) can be calculated as:

\begin{eqnarray}
%\xi = \dfrac{\rho g D}{2 \gamma}
R_{ST} = \dfrac{2 \gamma \xi}{\rho g D}
\label{eq:four}.
\end{eqnarray}

\begin{figure*}
 \includegraphics[width=\linewidth]{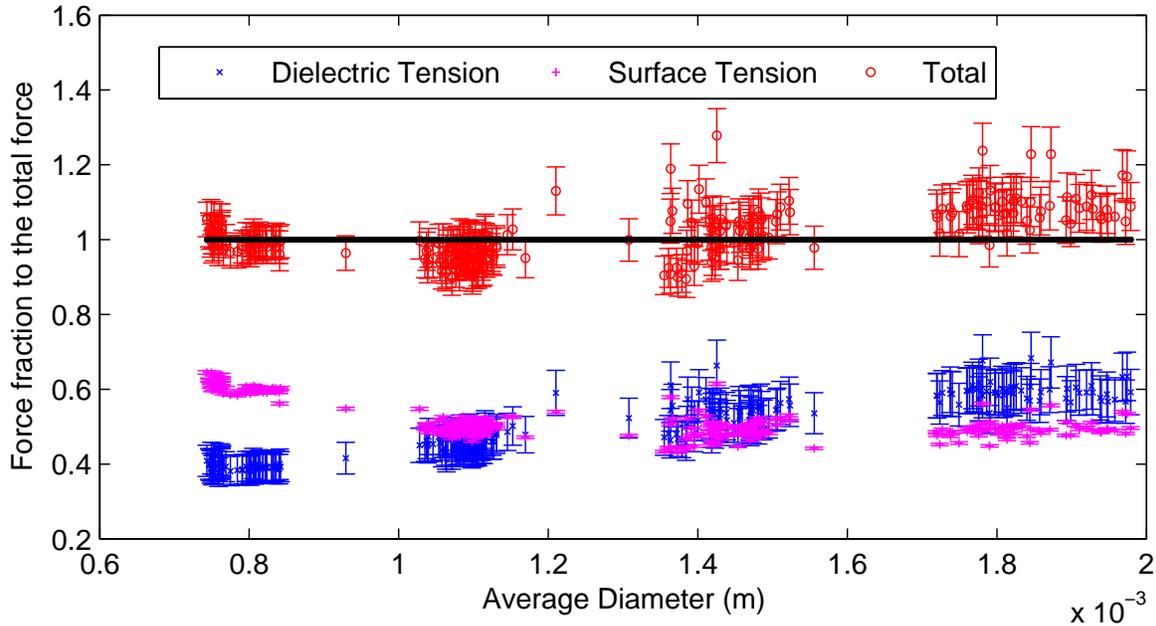}
 \caption{\label{fig:comparison}Ratio of the calculated forces to the force needed to hold the bridge.
 This result suggests that a sum of the surface tension and dielectric tension are convenient
 to explain the vertical equilibrium of the bridge. $l = 14mm$.}
 \end{figure*}

We have calculated $R_{DE}$ and $R_{ST}$ and also the sum of the two fractions for our experimental
data as plotted in figure \ref{fig:comparison}. The results suggest that the sum of the dielectric
force as calculated by Widom et al \cite{widom_t} and the surface tension force as calculated by Aerov \cite{aerov_t}
are sufficient for the vertical equilibrium of the bridge. Also the two forces have the same
order of magnitude, both being important and neither of the forces are negligible.

\section{\label{sec:Conc}Conclusion}
We have experimentally investigated the forces holding the floating water bridge against gravity.
By analysing the shape of the bridge from top and side views and evaluating the electric field
using a numerical simulation, we have estimated the forces of dielectric tension and surface tension
and compared them to the weight of the bridge.

Our results show that the vertical components of the two forces of 
dielectric tension in water and surface tension hold the
bridge against gravity. Our data shows that in smaller diameters of the bridge
the effect of surface
tension gets more important, while in thick bridges the dielectric tension is more 
important in holding the bridge. Our data shows that neither of the two forces are
negligible, each being responsible for about half of the weight of the bridge
and the sum of them is equal to the weight of the bridge.

We have shown that increasing the electric voltage of the power supply does not necessarily increase
the electric field along the bridge, because the cross-sectional area of the bridge varies 
fairly linear with current intensity. We suggest this to be a reason for not achieving bridges
longer than $2.5mm$ in high electric voltages in experiments, and also an explanation for
Aerov's claim \cite{aerov_t} about the dielectric tension hypothesis: 
"The electrostatic field hypothesis of the
bridge tension ($\tau \sim E^2$) is not really consistent
with experiments, because it allows the existence of bridges
longer than 4 cm in stronger fields, which seems to be not
the case."

We have shown that the stability of the floating water bridge can be fully explained with
the two forces of dielectric tension and surface tension. Changes in the structure of water
are not needed for explaining the stability.

\begin{acknowledgments}
We wish to thank Prof. K. Morawetz
and Prof. A. Aerov for helpful discussions in different stages of our investigation.
We acknowledge the International Young Physicists' Tournament (IYPT) society for
introducing this phenomenon and for discussions at IYPT 2012 Bad Saulgau Germany.
We also thank Sharif Applied Physics Research Center for it's financial support at
Medical Laser Physics Lab.
\end{acknowledgments}

\appendix*
\section{Upward Dielectric Force in the Water Bridge}
We directly calculate the vertical force
exerted to the dielectric according to the vertical gradients of the electric field. Note
that equipotential surfaces along the bridge must be normal to the surface of the bridge since
no electric current can flow normal to the surface. Thus equipotential surfaces are approximately
normal to the bridge centreline. This assumption is verified in numerical simulations as shown
in figure.\ref{fig:num1} a). As a result, equipotential surfaces are closer at the top of the bridge and
far at the bottom, so a higher electric field exists at the top. This causes an upward body
force to the bridge as:
\begin{eqnarray}
F_b = \dfrac{1}{2} \epsilon \nabla E^2
\label{eq:twoo}.
\end{eqnarray}
Assuming the equipotential surfaces to be flat and normal to the bridge centreline (as shown in Fig. \ref{fig:num1} a), the electric
field in the centre of the water bridge is totally in the direction of the length of the bridge
and can be calculated as a function of elevation $z$:
\begin{eqnarray}
E(z) = E_m \frac{1}{1-z\xi}
\label{eq:one_ap1}.
\end{eqnarray}
Where $E_m$ is the electric field at the altitude of zero which is the centreline of the bridge.
Thus at the centre of the bridge:
\begin{eqnarray}
\frac{\partial E^2}{\partial z} = 2 E_m ^2 \xi
F_b = \epsilon E^2 \xi
\label{eq:two_ap1}.
\end{eqnarray}
This body force must equal to the gravitational body force in case of equilibrium,
thus:
\begin{eqnarray}
\epsilon E^2 \xi = \rho g
\Rightarrow \xi = \frac{\rho g}{\epsilon E^2}
\label{eq:four_ap1}.
\end{eqnarray}
Which is the same as equation \ref{eq:two} derived from the theory of Widom et al \cite{widom_t}.
This might be a reason showing the existence and correctness
of the dielectric tension calculated therein.
%\end{verbatim}

\bibliography{Namin_waterbridge1}% Produces the bibliography via BibTeX.

\end{document}